\documentclass[aps, prb, reprint, superscriptaddress, amsmath, amssymb, letterpaper, showpacs]{revtex4-1}
\usepackage{graphicx}
\usepackage{siunitx}

\begin{document}

\title{Variable range hopping theory for the nodal gap in strongly underdoped  cuprate Bi$_2$Sr$_{2-x}$La$_x$CuO$_{6+\delta}$}
\author{Wei-Qiang Chen}
\affiliation{Department of Physics, South University of Science and Technology of China, Shenzhen, China}
\author{Jun-Qiu Zhang}
\affiliation{Department of Physics, South University of Science and Technology of China, Shenzhen, China}
\author{T. M. Rice}
\affiliation{Institut f\"{u}r Theoretische Physik, ETH Z\"{u}rich, CH-8093, Z\"{u}rich, Switzerland}
\author{Fu-Chun Zhang}
\affiliation{Department of Physics,  Zhejiang University, Hangzhou, China}
\affiliation{Collaborative Innovation Center of Advanced Microstructures, Nanjing, China}

\begin{abstract}
  Recent angle resolved photoemission spectrascope (ARPES) experiments on strongly underdoped Bi$_2$Sr$_{2-x}$La$_x$CuO$_{6+\delta}$ cuprates have reported an
  unusual gap in the nodal direction. Transport experiments on these cuprates found variable range hopping behavior
  observed. These cuprates have both electron and hole doping which has led to proposals that this cuprate is analogous
  to a partially compensated semiconductors.  The nodal gap then corresponds to the Efros-Schklovskii(ES) gap in such
  semiconductors.  We calculate the doping dependence and temperature dependence of a ES gap model and find support for
  an Efros-Schklovskii model.
\end{abstract}

\maketitle

The parent compound of high temperature cuprate superconductor is a Mott insulator with antiferromagnetic(AFM) long
range order \cite{review}.  With doping of holes into the system, the AFM order is suppressed at first and finally
vanishes at hole concentration $\sim 3\% - 5\%$.  When the hole concentration exceeds a critical doping $x_{c}$, the
system becomes superconducting.  The evolution of the system with hole concentrations is one of the central issues in
the field.  Many experiments have been performed in the very underdoped and low temperature regime and various phenomena
have been observed include spin glass, spin fluctuations, charge perturbations etc \cite{review}.  One problem is that
the impurity effect in this regime is strong because of the low carrier density and high impurity concentration (each
impurity introduces one hole).  This has led to debates about which phenomena are intrinsic or extrinsic due to the
impurities.  So it is timely to investigate the impurity effects in this regime.  Recently, Yingying Peng et
al. performed an ARPES and transport experiments on a series of Bi$_2$Sr$_{2-x}$La$_x$CuO$_{6+\delta}$ samples in this
regime.  They observed the variable range hopping behavior in the transport experiments and an energy gap along nodal
direction in the ARPES spectra\cite{exp1}.  By analogy to a partially compensated semiconductor, we attribute the energy
gap to the soft gap opened in the impurity band due to the Coulomb interaction between the localized electrons.  This
indicates that the low energy physics in this material is dominated by impurity effects.  Further experimental
investigations of these materials will help to separate the extrinsic phenomena originating from impurity effects from
the intrinsic phenomena.

\begin{figure}[htbp]
\centerline{\includegraphics[width=0.3\textwidth]{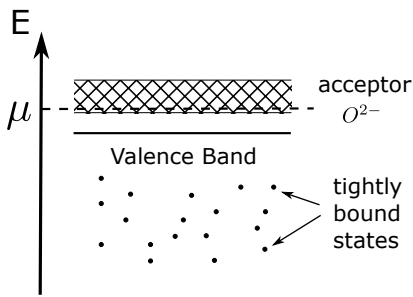}}
\caption[]{\label{fig:sketch} A schematic figure of the band structure in lightly doped Mott insulators, Bi$_2$Sr$_{2-x}$La$_x$CuO$_{6+\delta}$ with both hole and electron dopants, similar to partially compensated semiconductors. The upper Hubbard band(not shown) acts as a conduction band, and the valence band is the lower Hubbard(or Zhang-Rice) singlet band.
$La^{3+}$ impurities form a donor band below the bottom of the conduction band, and the compensating $O^{2-}$ ions form an acceptor band above the top of the valence band.  Some holes and electrons form local bound states due to strong attraction between nearby $La^{3+}$ and extra $O^{2-}$ ions, with energies well below the top of the valence band.  The chemical potential lies within the acceptor band due to the excess of holes.}
\end{figure}

In this paper, we examine the results of a recent series of ARPES and transport experiments on a series of
Bi$_2$Sr$_{2-x}$La$_x$CuO$_{6+\delta}$ samples with low hole doping, lying just below the critical doping for the onset
of superconductivity.  These cuprates are unusual in that large concentrations of La donors were added, which are
compensated by O acceptors. A series of samples were prepared with a large La donor concentration of $x=0.84$ which were
annealed in an O atmosphere. The result was samples whose net carrier concentration, P ,in the CuO$_2$ plane is
$P<0.10$.  Superconductivity was observed in samples at the upper end of the doping range. Low hole doping in the planar
CuO$_2$ can only be achieved in this way, when La donors and O acceptors are arranged in tightly bound neutral states
leaving a small number of remaining holes in the more extended states in the CuO$_2$ planes. The values of the net
carrier densities that result was estimated from a series of experiments, e.g resistivity, Hall coefficient and
thermoelectric power. This procedure led to estimates of the critical concentration for superconductivity of $P =0.1$.
The spatial distribution of these different states is shown schematically in Fig \ref{fig:sketch}. We will concentrate
on the nonsuperconducting samples with net hole concentrations $P < 0.1$. We simulate the density of states with varying
energy and temperature and compare with the results of the ARPES experiments.

In the ARPES experiments, one observes a peak-dip-hump structure.  In the case with hole doping $0.03 < P < 0.1$, a gap
is observed below the peak.  The magnitude of the gap on the underlying Fermi surface is found to be k-dependent,
minimal at the nodal direction and maximal in antinodal region.  The gap value is reduced with increasing hole doping
and vanishes at $P \sim 0.10$ where the system undergoes an insulator-superconductor transition.  With increasing the
temperature, the peak becomes weaker and weaker.  Another very important phenomena observed in transport experiments on
samples in this doping range, namely variable range hopping (VRH) behavior in the resistivity \cite{exp1,exp2}.  VRH was
proposed by Mott and has been well studied in semiconductors with localized impurity states forming impurity
bands.\cite{MOTT} In a partially compensated semiconductors, both acceptors and donors are present which form the
acceptor bands and donor bands respectively.  In a hole doped case, the donor bands are completely empty, while the
acceptor bands are partially filled.  At low temperatures, the hopping of holes can only happen between acceptor states
which lie very close to the chemical potential.  Because the localised acceptors are randomly distributed, acceptors
with a small energy difference are well separated.  So the characteristic hopping length of holes will increase with
decreasing temperature, leading to VRH.  VRH makes the system an insulator though the acceptor band is only partially
filled, leading to a resistivity which varies as $\rho_0 \exp[(T_0/T)^{\alpha}]$ with temperature.  The exponent
$\alpha$ is $1/3$ in 2D if there is a finite DOS at the Fermi level \cite{MOTT}.  But Efros and Schklovskii (ES) pointed
out later that because of the long range Coulomb interaction between the localised charges, a soft gap will open at
chemical potential at zero temperature\cite{ES}.  The direct consequence of the opening of ES gap is that the exponent
becomes $\alpha = 1/2$ at very low temperature.  With increasing temperature, the ES gap is filled and the exponent
$\alpha$ will be reduced to $1/3$.

In the experiments on Bi$_2$Sr$_{2-x}$La$_x$CuO$_{6+\delta}$, the transport data was fitted with the VRH form for
both $\alpha = 1/3$ and $1/2$.  But the fitting to an exponent $1/2$ is only good at low temperature, while the fitting to $1/3$ is
good at much larger temperature region \cite{FITTING}.  On one hand, this indicates a VRH physics in the
Bi$_2$Sr$_{2-x}$La$_x$CuO$_{6+\delta}$ system.  On the other hand, the exponent indicate a gap opens at low temperature
which fills up at the temperature is raised.  [Similar with the partially compensated semiconductor, the
Bi$_2$Sr$_{2-x}$La$_x$CuO$_{6+\delta}$ material also has two kinds of dopants, the La$^{3+}$ and the extra O$^{2-}$
which introduce the electrons and holes respectively.  So it is natural to make an analogy between the
Bi$_2$Sr$_{2-x}$La$_x$CuO$_{6+\delta}$ and a partially compensate semiconductors, where the La$^{3+}$ and extra O$^{2-}$
play the role of donors and acceptors respectively.]

As discussed above, the VRH behavior in Bi$_2$Sr$_{2-x}$La$_x$CuO$_{6+\delta}$ material at low temperature indicate the
opening of an ES gap.  This gap should correspond to the gap observed in the experiments \cite{exp1, exp2}. To investigate the
properties of the ES gap, we follow earlier calculations\cite{ES,BEGS,DLR} and study the following classical
Hamiltonian
\begin{align}
\label{eq:4}
H & = \sum_i n_i \phi_i + \frac{V}{2} \sum_{i \neq j} \frac{n_i n_j}{\mathbf{r}_{ij}},
\end{align}
where $i$ is the index of the centre of an impurity in the acceptor bands, $\phi_i$ is the energy of the
corresponding localised state, $n_i = 0, 1$ is the occupation number of holes in the state,
$V = e^2 / 4 \pi \epsilon_r \epsilon_0 a$ is the coupling strength of the Coulomb repulsion between two holes,
$\epsilon_r$ is the dielectric constant, $\mathbf{r}_{ij}$ is the distance between two impurities i and j.  In our
calculation, we assume that the impurities form a square lattice for simplicity.  This approximation will not affect the
result because only the states around the chemical potential are important, and those states are spatially separated and
have randomly distributed energies as pointed out in ref. \cite{DLR}.  As mentioned above, most O acceptors form
tightly bound neutral states with La donors and only a small fraction of extra O$^{2-}$ show up in the acceptor band,
i.e. the effective concentration $n_a$ of  O acceptors per Cu-site is much smaller than the nominal one
$\delta = (x+P)/2 = 0.42 + P/2$.  This leads to the uncertainty to estimate the charge carrier's filling $\nu_h$ in the
acceptor band, which is defined by
\begin{align}
\label{eq:3}
\nu_h = P / n_a,
\end{align}
with $P$ the concentration of doped holes in a Cu-oxide plane.  Since $n_a < \delta$, we have $\nu_h > P/\delta$.
While the precise relation between $n_a$ and $\delta$, hence the relation between $\nu_h$ and $P$, depends on the
specific material, it is reasonable to assume $\nu_h$ and $P$ are monotonic: a larger value of $\nu_h$ corresponds to a
larger $P$ in ARPES. In the calculations below, we consider three different values of $\nu_h$: 0.88, 0.92, and 0.96.

 First, we consider the density of states(DOS) at zero temperature of a 10$\times$ 10 lattice with periodic boundary
conditions.  The DOS are calculated and averaged over 10$^6$ impurity configurations.  In the calculations, the
bandwidth of the acceptor band is chosen to be 200 \si{meV}, i.e. $\phi_i$ is generated randomly in the interval
$[-100, 100] \si{meV}$.  The Coulomb coupling strength $V$ is chosen to $100 meV$ which corresponds to
$\epsilon_r \approx 38$.  The electron distribution with lowest energy for each configuration is achieved with the
procedure used in Ref. \onlinecite{BEGS} and \onlinecite{DLR}.  At first, we generate an initial distribution of the
holes randomly for a given impurity configuration.  Then we apply the so-called $\mu$-sub procedure.  We calculated the
single particle energy for each state with
\begin{align}
\label{eq:1}
E_i & = \phi_i + \sum_{j \ne i} \frac{n_{j}}{r_{ij}},
\end{align}
where $r_{ij}$ is the shortest distance between two sites $i$ and $j$.  If $E_i$ of highest occupied site is larger
than $E_i$ of the lowest empty site, we will move the hole from the highest occupied site to lowest empty site.  Then
we recalculate $E_i$ and do the check again and again until all $E_i$ of occupied sites are less than any
$E_i$ of the empty sites.

\begin{figure}[htbp]
\centerline{\includegraphics[width=0.4\textwidth]{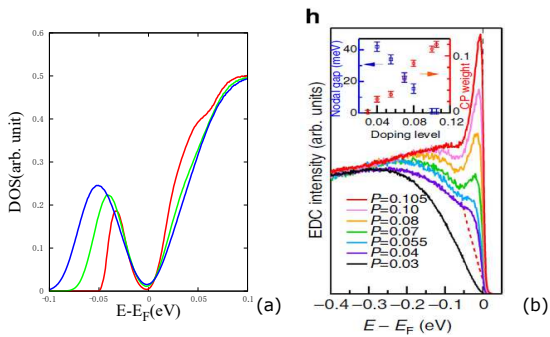}}
\caption[]{\label{fig:dos} (a) Calculated electron density of states (DOS) for various of filling factors of holes,
  $\nu_h$ in the impurity acceptor band: $\nu_h = 0.88$(blue line), $\nu_h=0.92$(green lines), and $\nu_h=0.94$(red
  lines), respectively. (b) The ARPES data on energy distribution curve (EDC) for various hole dopings $p$, from fig. 3
  in Ref. \onlinecite{exp1}. The right panel in (b) shows only the occupied electron states which are observed in
  ARPES. Note that the position of the DOS peak shown in (a) shifts towards zero as $\nu_h$ increases, qualitatively
  consistent with the ARPES data, showing the peak position in EDC shifts towards zero as $p$ hence $\nu_h$ increases.}
\end{figure}

In the next stage, we check the energy for a single hole, which hops with
\begin{align}
\label{eq:2}
E_{ji} = E_j - E_i - \frac{1}{r_{ij}},
\end{align}
where $i$ is an occupied site and $j$ is an empty site.  The hop will be performed if $E_{ji} < 0$.  After each hop, we
redo the $\mu$-sub process and check all possible single hole hop until all $E_{ji}$ are positive.  Then
we take the resulting hole distribution as one candidate ground state.  After 5,000 iterations of the whole process
for a given impurity configuration, we choose the one with lowest energy from all the candidates and choose it as the
ground state for that impurity configuration.

With the single particle energy of the ground states, we can get the DOS by averaging over all the impurity
configurations, the result is shown in fig.~\ref{fig:dos}(a), where the DOS at $\nu_h = 0.88$(blue line), 0.92(green
line) and 0.96(red line) are depicted. To compare with the ARPES experiment, the DOS is presented in electron notation.
It is obvious that a soft gap opened at the chemical potential for all of the three cases at low temperatures.  The
finite DOS at chemical potential at larger hole doping case may due to the finite size effect.  The most important
feature is that the peak position is shifted towards zero as the hole doping $p$ increases, which is qualitatively
consistent with the experimental result shown in fig.~\ref{fig:dos}(b), which is extracted from fig. 3 in
Ref. \onlinecite{exp1}.

\begin{figure}[htbp]
\centerline{\includegraphics[width=0.4\textwidth]{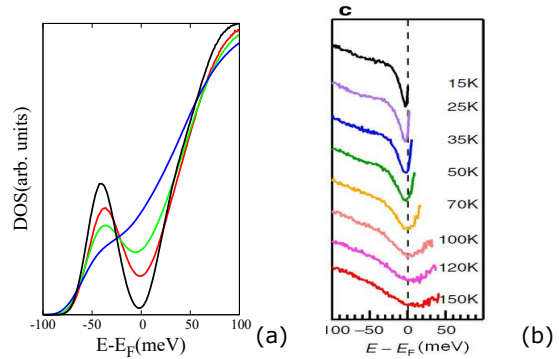}}
\caption[]{\label{fig:dos_t} (a) Calculated DOS at impurity acceptor band filling factor $\nu_h = 0.92$ at various
  temperatures: $k_BT = 0$(black line), $6$meV(red line), $10$meV(green line), and $20$meV(blue line), respectively. (b)
  The ARPES result for energy distribution curve of at various temperatures, extract from fig. 4 in
  Ref. \onlinecite{exp1}. See the text for a discussion of the qualitative agreement between the theory and
  experiment. }
\end{figure}

Then we study the DOS at finite temperature with the Monte Carlo technique used in ref.\cite{DLR}.  At first, we
generate 10$^4$ random configurations of $\phi$.  For each configuration, we calculate the single particle energies from
higher temperature to lower temperature with a standard Monte Carlo algorithm.  At each temperature, we drop the first
10$^4$ hole configurations to remove the memory of higher temperature and keep the single particle energies for the
next 10$^5$ hole configurations.  By averaging over all the impurity configurations, we get the DOS at finite
temperature.  The result for $\nu_h = 0.92$ at temperatures $k_BT = 0$ (black line), 6meV (red line), 10meV(green line)
and 20meV(blue line) are depicted in fig.~\ref{fig:dos_t}(a).  With increasing temperature, the amplitude of the peak
decreases while the DOS at the chemical potential increases leading to the closing of the ES gap.  Note the position of the
peaks shift only slightly to lower energy with increasing temperature.  These results are consistent with the experiment
shown in fig.~\ref{fig:dos_t}(b) extracted from fig. 4 in Ref. \onlinecite{exp1}.

Another feature of the ARPES experiments is the momentum dependence of the ES gap which is minimal at the nodal point and
increases away from the nodal point.  This momentum dependence may originate from the spatial distribution of
the impurities which could be affected by various effects like clustering etc. at such high impurity densities.  But in
our model, we only consider the random distributed impurities because of the lack of information on the actual spatial
distribution.

We have carried out simulations of the electron density of states for lightly doped cuprates with strong random
potentials and compensated electron and hole dopings and compared them to the results of ARPES experiments on
nonsuperconducting Bi$_2$Sr$_{2-x}$La$_x$CuO$_{6+\delta}$ samples whose transport properties show variable range hopping
at low temperatures. The ARPES experiments measure only states occupied by electron with energies strictly below the
chemical potential at low temperatures, extending to slightly above as T is raised. Our simulations show good agreement
with the key features in the density and temperature dependence of the electron density of states and support the
interpretation put forward by X. J. Zhou and collaborators of their experimental results.

\begin{acknowledgments}
  We thank X. J. Zhou for interesting discussions and allowing us to reuse the figures in their paper.  WQ Chen is
  partly supported by NSFC 11374135, and FC Zhang thanks NSFC 11274269, and National Basic Research Program of China
  (No. 2014CB921203).
\end{acknowledgments}

\end{document}